\documentclass[magazine]{IEEEtran}
\IEEEoverridecommandlockouts
\usepackage{cite}
\usepackage{amsmath,amssymb,amsfonts}
\usepackage{amssymb}
\usepackage{algpseudocode}
\usepackage{amsfonts}
\usepackage{graphicx}
\usepackage{fancyhdr}
\usepackage{cases}
\usepackage{textcomp}
\usepackage{extarrows}
\usepackage{algorithm,algpseudocode}
\usepackage{multirow}
\usepackage{makecell}
\usepackage{booktabs}
\usepackage{caption}
\usepackage{subcaption}
\usepackage{orcidlink}
\hypersetup{hidelinks} 

\usepackage{multirow,tabularx}
\usepackage{mathtools}
\usepackage{xcolor}
\usepackage[english]{babel}
\usepackage{amsthm}
\usepackage[left]{lineno}

\usepackage{caption}
\def\BibTeX{{\rm B\kern-.05em{\sc i\kern-.025em b}\kern-.08em
    T\kern-.1667em\lower.7ex\hbox{E}\kern-.125emX}}
    
\begin{document}
\bstctlcite{BSTcontrol}
\title{Robust Deep Learning-Based Physical Layer Communications: Strategies and Approaches}

\author{
	\IEEEauthorblockN{
	Fenghao Zhu$^{\orcidlink{0009-0006-5585-7302}}$,
	Xinquan Wang$^{\orcidlink{0009-0005-9986-7054}}$,
	Chen Zhu$^{\orcidlink{0009-0004-8676-1915}}$,
	Tierui Gong$^{\orcidlink{0000-0002-5136-3448}}$,
	Zhaohui Yang$^{\orcidlink{0000-0002-4475-589X}}$, \\
	Chongwen Huang$^{\orcidlink{0000-0001-8398-8437}}$, 
	Xiaoming Chen$^{\orcidlink{0000-0002-1818-2135}}$,
	Zhaoyang Zhang$^{\orcidlink{0000-0003-2346-6228}}$ and
	M\'{e}rouane~Debbah$^{\orcidlink{0000-0001-8941-8080}}$,~\IEEEmembership{Fellow,~IEEE}}

\thanks{The work was supported by the National Natural Science Foundation of China under Grant 62331023 and 62394292, Zhejiang Provincial Science and Technology Plan Project under Grant 2024C01033, Zhejiang University Global Partnership Fund, and  Fundamental Research Funds for the Central Universities under Grant No. 226-2024-00069. 
(\textit{Corresponding author: Chongwen Huang.})
\par
F. Zhu, X. Wang, Z. Yang, C. Huang, X. Chen and Z. Zhang are with the College of Information Science and Electronic Engineering, Zhejiang University, Hangzhou 310027, China, and Zhejiang Provincial Key Laboratory of Multi-Modal Communication Networks and Intelligent Information Processing, Hangzhou 310027, China. (E-mails:
\href{mailto:zjuzfh@zju.edu.cn}{\{zjuzfh}, 
\href{mailto:wangxinquan@zju.edu.cn}{wangxinquan}, 
\href{mailto:yang_zhaohui@zju.edu.cn}{yang\_zhaohui},
\href{mailto:chongwenhuang@zju.edu.cn}{chongwenhuang},
\href{mailto:chen\_xiaoming@zju.edu.cn}{chen\_xiaoming},
\href{mailto:zhzy@zju.edu.cn}{zhzy}\}@zju.edu.cn).}

\thanks{C. Zhu is with Polytechnic Institute, Zhejiang University, Hangzhou 310015, China. (E-mail: \href{mailto:zhuc@zju.edu.cn}{zhuc@zju.edu.cn}).}

\thanks{T. Gong is with Nanyang Technological University, Singapore 639798 (E-mail: \href{mailto:trgTerry1113@gmail.com}{trgTerry1113@gmail.com}).}

\thanks{M. Debbah is with KU 6G Research Center, Department of Computer and Information Engineering, Khalifa University, Abu Dhabi 127788, UAE (E-mail: \href{mailto:merouane.debbah@ku.ac.ae}{merouane.debbah@ku.ac.ae}).}
}
\maketitle

\pagestyle{empty}  
\thispagestyle{empty} 

\begin{abstract}
Deep learning (DL) has emerged as a transformative technology with immense potential to reshape the sixth-generation (6G) wireless communication network. By utilizing advanced algorithms for feature extraction and pattern recognition, DL provides unprecedented capabilities in optimizing the network efficiency and performance, particularly in physical layer communications. Although DL technologies present the great potential, they also face significant challenges related to the robustness, which are expected to intensify in the complex and demanding 6G environment. Specifically, current DL models typically exhibit substantial performance degradation in dynamic environments with time-varying channels, interference of noise and different scenarios, which affect their effectiveness in diverse real-world applications. This paper provides a comprehensive overview of strategies and approaches for robust DL-based methods in physical layer communications. First we introduce the key challenges that current DL models face. Then we delve into a detailed examination of DL approaches specifically tailored to enhance robustness in 6G, which are classified into data-driven and model-driven strategies. Finally, we verify the effectiveness of these methods by case studies and outline future research directions.
\end{abstract}

\section{Introduction}\label{sec:intro}
The sixth-generation (6G) wireless communication network is envisioned to revolutionize the telecommunications landscape by incorporating a wide range of advanced technologies, particularly in the physical layer, such as new modulation schemes and extremely large antenna arrays. These advances aim to deliver enriched and immersive experiences, provide ubiquitous and seamless coverage, and facilitate innovative forms of collaborations \cite{9390169}. Even if these technologies improve the system capacity and throughput significantly, they often involve substantial algorithmic complexity due to the handling of high dimensional channel matrices, especially in complex and dynamic environments. To mitigate this challenge, advanced technologies, such as deep learning (DL), have been developed to reduce the associated overhead \cite{Jakob_deep}. Integrating DL into wireless communication systems facilitates better signal quality, reduced interferences, and improved energy efficiency. The synergy between DL and 6G communications is anticipated to drive unprecedented connectivity and intelligent communication solutions with significantly enhanced performance and efficiency.
\par
Despite the great potential of DL technologies in physical layer communications, ensuring the practical viability and reliability of these methods requires addressing significant inherent challenges \cite{gmml}. A primary concern revolves around the robustness of DL models. In the context of this paper, robustness refers specifically to the ability of a DL based communication system to maintain consistent and reliable performance despite variations and uncertainties encountered in real-world wireless environments. These variations may include time-varying channel conditions, the presence of diverse noise and interference types beyond standard assumptions like additive white Gaussian noise, and operation across different deployment scenarios or hardware configurations potentially unseen during the initial training phase.
\par
In practice, evaluating such robust DL models often relies on standard physical layer performance metrics, such as bit error rate (BER), spectral efficiency (SE) and so on, which provide consistent benchmarks. A key feature of a robust DL model is its ability to maintain desirable performance levels (e.g., consistently low BER for reliability and high SE for efficiency) across these diverse and challenging conditions. Consequently, a robust DL model should exhibit adaptability to unseen or dynamically changing conditions, alongside computational stability suitable for real-time processing demands.

\renewcommand{\arraystretch}{1.5}
\begin{table*}[t]
\caption{Summary of Strategies and Approaches for DL-Based Robust Physical Layer Communications}
\centering
\begin{tabular}{|c|c|m{7cm}|c|c|c|}
\hline
\textbf{Strategies} & \textbf{Approaches} & \makecell{\textbf{Feature}} & \multicolumn{3}{c|}{\textbf{Addressed Issues}} \\ \cline{4-6}
&  & \multicolumn{1}{c|}{} & \raisebox{-1mm}{\makecell{\textbf{Time-Varying}\\ \textbf{Channels}}} & \raisebox{-1mm}{\makecell{\textbf{Interference} \\\textbf{of Noise}}} & \raisebox{-1mm}{\makecell{\textbf{Different} \\ \textbf{Scenarios}}} \\ \hline
\multirow{3}{*}{\makecell{Data-\\Driven}} 
& \makecell{Transfer \\ Learning \cite{MAML1}} & Utilizes pre-trained models on large datasets for fine-tuning with smaller, task-specific data. & & & \makecell{\checkmark} \\ \cline{2-6} 
& \makecell{Data-Driven\\ Meta Learning \cite{meta_deepsic} }& Leverages training data division to improve the fine-tuning adaptation speed. & \makecell{\checkmark} & & \makecell{\checkmark} \\ \cline{2-6} 
& \makecell{Data-Driven\\Hybrid Learning \cite{zhu2023robust}} & Combines cross-scenario data sources with a self-adaptive proportion for learning. & & & \makecell{\checkmark} \\ \hline
\multirow{3}{*}{\makecell{Model-\\Driven}} 
& Deep Unfolding \cite{TAI_Unfolding} & Unfolds an iterative algorithm into a neural network by representing each iteration as a network layer. & & \makecell{\checkmark} & \\ \cline{2-6} 
& \makecell{Model-Driven \\Meta Learning \cite{gmml} }& Optimizes the search space trajectory within the optimization space. & \makecell{\checkmark} & \makecell{\checkmark} & \\ \cline{2-6} 
& \makecell{ODE-Based\\ Learning \cite{ncps_beamforming}} & Leverages ODE-based continuous and adaptive neurons to model physical and dynamic communication processes. & \makecell{\checkmark} & \makecell{\checkmark} & \\ \hline
\end{tabular}
\vspace{-3mm}
\label{tab:summary}
\end{table*}

\par
The remainder of this article is organized as follows: Section \ref{sec:challenges} outlines the challenges that current DL models face. Section \ref{sec:robust_schemes} explores robust DL solutions for future wireless networks. Section \ref{sec:cases} presents case studies validating the effectiveness of these robust methods. Potential future research directions are provided in Section \ref{sec:robust_challenges}, and the conclusions are lastly made in Section \ref{sec:conclusion}.

\section{Challenges}\label{sec:challenges}
In this section, we explore the primary obstacles encountered when deploying DL techniques in complex and dynamic wireless physical layer. These challenges are detailed below.

\subsubsection{Time-Varying Channels}\label{Time-Varying Channels}
Wireless channels are highly dynamic and vary rapidly due to factors, such as user movements, obstacles, and weather changes. For instance, a user moving through an urban environment may experience different levels of signal obstruction from buildings, vehicles, and other structures. The time variability of wireless channels requires DL models capable of adapting to dynamic changes in real-time, which can be computationally heavy for efficient implementations. Maintaining satisfactory performance in the context of fluctuating conditions represents a significant challenge for DL models. 
This challenge significantly intensifies in 6G. Higher frequencies like millimeter-wave (mmWave) or Terahertz increase channel sensitivity to blockage and atmospheric effects. Furthermore, the dynamic control technology like reconfigurable intelligent surfaces would introduce additional rapid channel dynamics. Both factors necessitate exceptional real-time adaptability from DL models.

\subsubsection{\textit{Interference of Noise}}\label{Interference of Noise}
Noise is pervasive in wireless communications, where transmission signals are susceptible to degradation by background noise, significantly impairing the reliability and accuracy of communication systems. These noises can lead to distorted signal, erroneous data transmissions, and reduced data throughput, ultimately compromising the overall performance of wireless communication networks. For instance, inter-user interference and background noise can distort the received pilot signal, making it challenging for DL models to accurately recover the channel state information (CSI). In 6G, this issue becomes more complex. The anticipated massive connectivity and ultra-dense deployments will likely lead to more intricate interference patterns. As a result, the development of robust DL models that can effectively mitigate the impact of noise interference is crucial for reliable and efficient communication systems.

\subsubsection{Different Scenarios}\label{Different Scenarios}In the era of ubiquitous communications, the scope of communication services is expanding, characterized by continuous growth in modalities and increasingly flexible and dynamic scenarios. However, the complex data distributions and diverse operation modes in such dynamic and unpredictable scenarios impose remarkable challenges to the current DL models. The lack of out-of-distribution generalization ability of existing DL models can lead to a substantial performance degradation, making it difficult to ensure reliable operations \cite{gmml}. Specifically, a DL model trained under a certain configuration (e.g., user weights) may fail in the real-world deployments, where the conditions are diverse. This challenge is amplified in 6G, which envisions an unprecedented heterogeneity of applications, diverse device types, and deployment environments. This increases the design complexity of DL models, as they must accommodate a wide range of communication requirements from various scenarios, and a multitude of hardware and software configurations across user devices and base stations. Retaining the flexibility and ensuring reliable operation of DL models across these diverse and dynamic communication scenarios is of particular interest.

\section{DL-Based Robust Schemes}\label{sec:robust_schemes}
\begin{figure*}[t]
	\begin{center}
		\centerline{\includegraphics[width=0.90\linewidth]{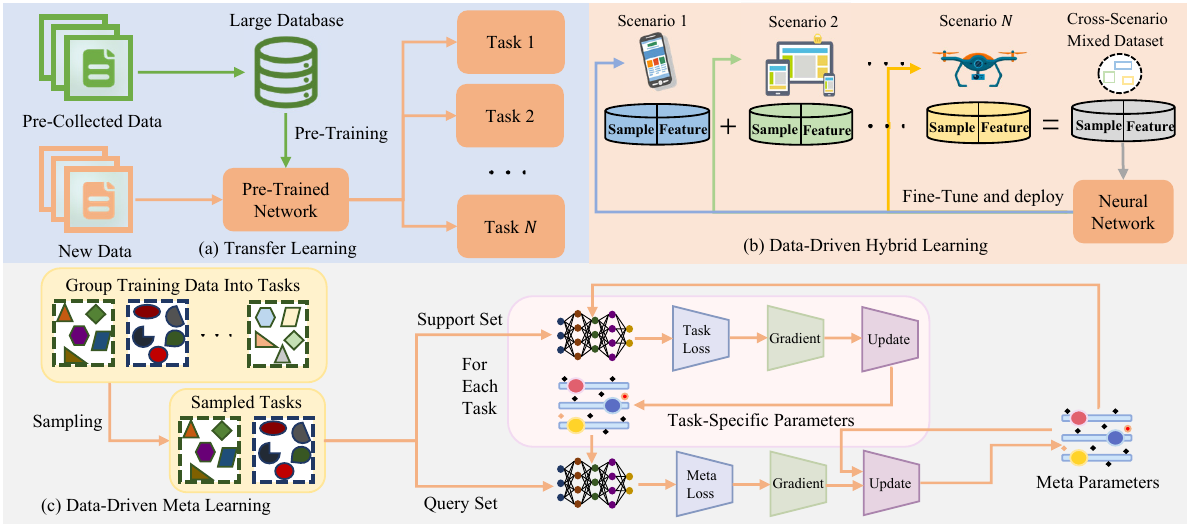}}  \vspace{-0mm}
		\captionsetup{font=footnotesize, name={Fig.}, labelsep=period} 
		\caption{\, Data-driven methods for DL-based physical layer communications. (a) Transfer Learning leverages models pre-trained on large datasets for fine-tuning on specific tasks. (b) Data-Driven Hybrid Learning enhances robustness by training on mixed datasets from diverse scenarios. (c) Data-Driven Meta Learning enables rapid adaptation to new tasks by learning optimal initial parameters from sampled tasks using support and query sets.}
		\label{data_driven}
		\vspace{-3mm}
	\end{center}
\end{figure*}

\begin{figure*}[t]
	\begin{center}
		\centerline{\includegraphics[width=0.90\linewidth]{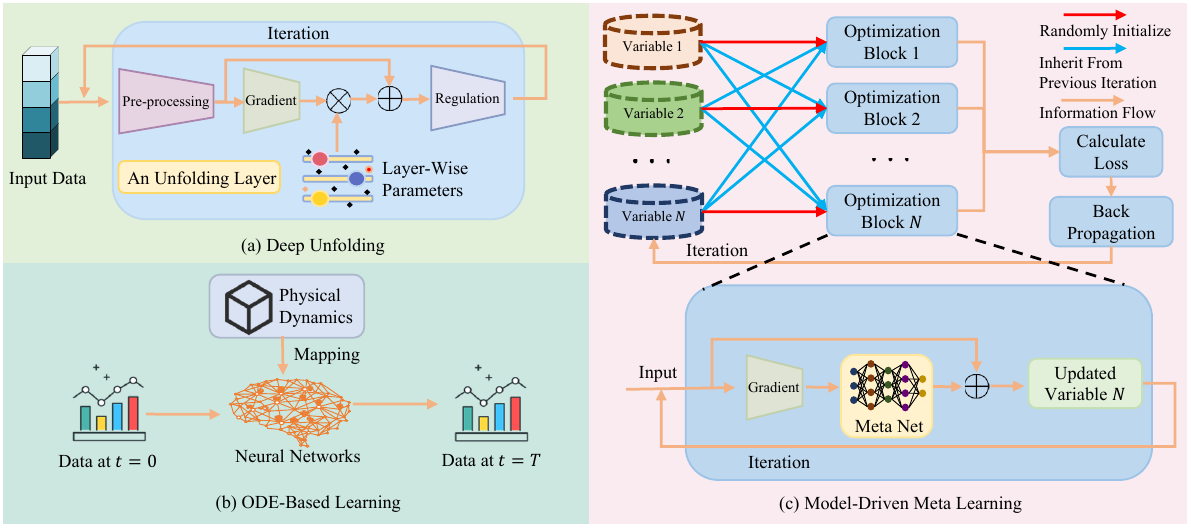}}  
		\vspace{-0mm}
		\captionsetup{font=footnotesize, name={Fig.}, labelsep=period} 
		\caption{\, Model-driven methods for DL-based physical layer communications. (a) Deep Unfolding incorporates domain knowledge by unfolding iterative algorithms into a neural network with learnable parameters, where each layer of neurons represents an iteration in iterative algorithms. (b) ODE-Based Learning embeds differential equations within neural networks to model continuous system dynamics over time. (c) Model-Driven Meta Learning optimizes the update strategy itself, often using a meta-network to guide the iterative optimization process for variables.}
		\label{model_driven}
		\vspace{-3mm}
	\end{center}
\end{figure*}

In this section, we provide an overview of robust DL-based schemes that facilitate future physical layer communications, categorizing them into two primary paradigms: data-driven and model-driven strategies. Data-driven approaches leverage large datasets to learn patterns and make predictions, whereas model-driven approaches rely on mathematical models and domain expertise to inform DL-based solutions. We will first introduce the supervised learning and its limitations, and then delve into the distinct characteristics, advantages, and applications of the robust DL-based schemes, highlighting their potential to enhance the performance and reliability of physical layer communications. The features and benefits are summarized in Table \ref{tab:summary}. The diagrams for these strategies are illustrated in Fig. \ref{data_driven} and Fig. \ref{model_driven}.

\subsection{Traditional Supervised Learning and Limitations}

Traditional supervised learning (SL) serves as the foundational DL approach, where algorithms learn a mapping function from input features to desired output labels by analyzing extensive datasets of pre-labeled input-output pairs. The core mechanism involves identifying statistical patterns and relationships within this labeled training data, enabling the model to generalize and make predictions on previously unseen inputs. Within the domain of physical layer communications, SL has been widely employed for various tasks, such as channel estimation and beamforming. However, standard SL lacks robustness in dynamic wireless environments due to key limitations. SL requires vast and high-quality labeled data representative of the deployment scenario. Acquiring such data for diverse wireless conditions (locations, interference, hardware) is often costly, time-consuming, or infeasible. This need arises from the core SL assumption that training and deployment data distributions are identically distributed. Wireless environments frequently violate this assumption due to changing channels, interference, mobility, or hardware variations. Consequently, SL models trained on one data distribution generalize poorly when encountering different distributions (domain or distribution shift). The statistical correlations learned during training become invalid, leading to significant performance degradation on out-of-distribution data.

\subsection{Robust Data-Driven Strategy}

\subsubsection{Transfer Learning}
Typically, extensive training data is required to achieve high performance and accuracy in various tasks, especially in environments with diverse and changing conditions. However, obtaining accurate and representative training data is often challenging, particularly for new or evolving scenarios. Traditional supervised learning methods struggle with these requirements, leading to degraded performance when models encounter unseen environments or shifts in user distribution. Transfer learning addresses these issues by allowing models pre-trained on large datasets from similar tasks to be fine-tuned with smaller, task-specific datasets. This approach reduces the burden of data collection and processing, while also speeding up the training process and improving the generalization capabilities of the model. For example, a model trained on urban cellular data of a city can be effectively adapted for use in another city with minimal additional training. By leveraging the knowledge acquired from previous experiences, transfer learning facilitates the deployment of models in new scenarios at a lower cost, thereby enhancing the cross-scenario robustness.
\par
The authors of \cite{MAML1} investigated the effectiveness of transfer learning in beamforming by employing a deep transfer learning approach to address the signal-to-interference-plus-noise ratio (SINR) balancing problem in multi-user multiple-input single-output (MISO) downlink systems. The study utilized a deep neural network pre-trained on a large synthetic dataset, which simulated diverse wireless conditions. This pre-trained model was then fine-tuned with a smaller dataset from a real-world environment. The results demonstrated substantial improvements in beamforming accuracy and spectral efficiency (SE), highlighting the ability of transfer learning to adapt models to new, unseen network environments. This approach effectively mitigates performance degradation caused by differences between training and testing conditions, showcasing the potential of transfer learning to enhance beamforming performance in practical scenarios where network conditions vary. Moreover, transfer learning helps mitigate the issue of data scarcity in specific scenarios, such as mmWave communications, where obtaining extensive CSI data is challenging \cite{MAML1}. By transferring knowledge from sub-6 GHz band models to mmWave band models, the adaptation process becomes more efficient, leveraging existing data to improve high-frequency band performance.

\subsubsection{Data-Driven Meta Learning}
Data-driven meta learning focuses on enhancing the adaptability and generalization of learning algorithms by leveraging data to inform model updates. One prominent approach within this category is model-agnostic meta learning (MAML). MAML is designed to find initial model parameters that can be quickly adapted to new tasks with minimal fine-tuning, making it particularly suitable for wireless networks environments where data distribution can change rapidly.
\par
The MAML algorithm involves two main phases: meta-training and meta-testing. During meta-training, the model is exposed to a range of tasks to learn a set of parameters that are broadly effective. Each task is splited into support sets and query sets. Specifically, MAML first initializes the model parameters and then performs a few gradient updates on the support set of each task. After these updates, the performance of the model is evaluated on the query set, and the objective is to minimize the loss across all tasks. This training process ensures that the model parameters are optimized to be close to a solution that can be quickly adapted with minimal additional training.
\par
In wireless resource allocation, data-driven meta-learning, particularly MAML, offers substantial benefits by facilitating rapid adaptation to fluctuating channel conditions and interference patterns \cite{meta_deepsic}. By training the model on a variety of scenarios, MAML ensures that the model starts from an initialization that is well-suited for quick adjustments to new, unseen conditions. This capability enhances the robustness and efficiency in dynamic wireless environments, enabling effective and agile responses to changing network conditions.

\subsubsection{Data-Driven Hybrid Learning} Current supervised learning algorithms for wireless communications face significant challenges in acquiring high-quality and accurate CSI, particularly in dynamic environments with moving users. Training models on data from a single source often leads to severe performance degradation when tested in different environments with varying data distributions. Consequently, developing models that can effectively handle multiple scenarios without relying on extensive labeled data remains a key challenge for the practical deployment of DL in this field.
\par
To address these issues, \cite{zhu2023robust} proposed a self-supervised hybrid learning method for communications across various scenarios. This approach trains DL models without the need for labeled data, utilizing mixed training datasets to enhance robustness and generalization. It explored the relationship between the proportion of mixed training datasets and performance on test datasets, validating the proposed empirical formula through experiments on two distinct datasets.
\par
The proposed hybrid learning scheme offers several advantages. Firstly, it maintains robustness within a single dataset by leveraging self-supervised learning techniques, which reduce reliance on labeled data and improve model adaptability. Secondly, it enhances cross-dataset performance, enabling the DL model to generalize effectively across different datasets and environments. This adaptability is crucial for dynamic wireless systems, where environmental conditions and user behaviors continuously change. Furthermore, the hybrid learning method optimizes SE in the targeted deployment scenario by balancing the proportion of mixed training data. This ensures that the DL model performs well not only within the training environment but also when deployed in diverse real-world scenarios.

\subsection{Robust Model-Driven Strategy}

\subsubsection{Deep Unfolding}
Deep unfolding is a model-driven approach that integrates the interpretability of traditional signal processing algorithms with the learning capability of deep neural networks (DNNs). This method involves unfolding an iterative algorithm into a neural network architecture, where each iteration is represented as a layer in the network. The parameters of these layers are learned from data, which allows the network to adaptively optimize the algorithm.  
\par
One significant benefit of deep unfolding is its ability to leverage prior domain knowledge. Since the underlying iterative algorithm is well-understood, the network can be designed to maintain the constraints and properties of the problem, ensuring physically plausible solutions. Additionally, deep unfolding networks typically require fewer training samples compared to purely data-driven approaches, as the structure of the iterative algorithm provides a strong inductive bias. Consequently, unfolding algorithms are less susceptible to input variance disturbances, demonstrating higher stability and robustness compared to pure DL methods.
\par
In the context of physical layer communications, deep unfolding offers significant advantages. Traditional iterative algorithms, such as the weighted minimum mean square error (WMMSE) algorithm, are computationally demanding and may not always converge swiftly. By transforming these algorithms into neural networks through deep unfolding, the convergence process is accelerated while maintaining the theoretical foundations of the original algorithms. This approach results in a more efficient and effective beamforming solution. A general deep unfolding method was introduced in \cite{TAI_Unfolding} to tackle latency and overhead issues in practical wireless communication scenarios. This method has demonstrated promising prospects, for example, multiple-input multiple-output (MIMO) detection benefits from reduced complexity by avoiding certain matrix inversion operations. Additionally, error correction schemes in end-to-end communications can achieve lower bit error rates using the projected gradient descent method. Overall, deep unfolding provides robust and efficient solutions for complex problems in future wireless networks by integrating the interpretability of traditional algorithms with the adaptability of neural networks, leading to enhanced stability and reliability.

\subsubsection{Model-Driven Meta Learning} 

Model-driven meta learning refers to a meta learning paradigm that leverages mathematical models. Unlike data-driven approaches, which rely heavily on extensive training datasets to learn and generalize optimization strategies, model-driven meta learning optimizes strategies directly within the optimization space, rather than focusing on the target variables. A model-driven meta learning approach, as proposed in \cite{MLAM}, operates without extensive pre-training. It employs a training-while-solving approach, continually updating the optimization trajectory, which consists of a limited number of steps, to minimize the loss function by the end of the process.
\par
This approach differs from traditional greedy methods, which optimize variables sequentially and are prone to being trapped in local optima, especially for complex problems like multi-user beamforming. In contrast, model-driven meta learning considers the entire optimization trajectory, enabling it to tackle NP-hard problems more effectively.
\par
Moreover, model-driven meta learning demonstrates robust performance in noisy environments typical of practical wireless settings. For instance, in reconfigurable intelligent surfaces aided beamforming scenarios, it can effectively filter out noise and reduce the impact of channel estimation errors \cite{gmml}. By optimizing the entire search space trajectory rather than focusing solely on individual variables, meta learning dynamically adjusts and selects superior optimization paths.
\par
Additionally, model-driven meta learning utilizes smaller-scale networks compared to traditional deep learning models, leading to significant gains in energy and time efficiency. This makes it a practical choice for real-time applications in modern wireless communication systems where computational resources may be limited.

\renewcommand{\arraystretch}{1.2}
\begin{table}[t]
\centering\caption{Comparation and Tradeoff between Data-Driven and Model-Driven Methods}
\begin{tabular}{|p{2cm}|p{2.8cm}|p{2.8cm}|}
\cline{1-3}
\textbf{Feature} & \textbf{Data-Driven Methods} & \textbf{Model-Driven Methods} \\
\cline{1-3}
Data Requirement & High & Low \\
\cline{1-3}
Interpretability & Low (black-box) & High\\
\cline{1-3}
Handling Complexity & Good (can learn complex patterns) & Potentially limited by model accuracy \\
\cline{1-3}
Overfitting Risk & High (especially with limited data) & Low \\
\cline{1-3}
Training Cost & High & Generally lower \\
\cline{1-3}
Inference Cost & Low & Can vary \\
\cline{1-3}
Adaptability & Good with sufficient new data & May require architectural changes \\
\cline{1-3}
Physical Plausibility & Not inherently guaranteed & Embedded in model structure \\
\cline{1-3}
Domain Expertise & Not inherently required & Often integrated into model \\
\cline{1-3}
Out of Distribution & May struggle with unseen data & May struggle with uncovered scenarios \\
\cline{1-3}
\end{tabular}
\vspace{-3mm}
\label{tab:tradeoff}
\end{table}

\subsubsection{ODE-Based Learning}
Ordinary differential equations (ODEs) describe the time-based evolution of systems, linking physical processes with temporal changes \cite{ODE}. Leveraging this, ODE-based neural networks employ continuous modeling instead of discrete layers, enhancing robustness and efficiency. This approach allows them to directly learn complex continuous dynamics governing physical phenomena, making them well-suited for temporal tasks such as mobile channel prediction \cite{xiaozhuoranODE}.
\par
Classical examples of ODE-based neural networks include continuous-time recurrent neural networks (CT-RNNs) and ODE-long short-term memory (LSTM) networks. CT-RNNs use ODEs to model sequences in continuous time, making them suitable for handling real-world challenges like irregular time intervals and varying sampling rates. On the other hand, ODE-LSTMs integrate continuous-time modeling into the LSTM framework, enhancing their ability to manage continuous dependencies and dynamics, bridging the gap between discrete time steps and continuous-time processes. ODE-LSTMs offer improved flexibility and adaptability in modeling complex temporal sequences compared to traditional LSTMs. However, both ODE-LSTMs and CT-RNNs face challenges, including high computational complexity, longer training times, and issues with interpretability and training stability \cite{ODE}.

\par
Recently, a novel type of ODE-based neural network, known as liquid neural networks (LNNs), has been developed from first principles to address the above shortcomings. LNNs fundamentally differ from other models in their neuron operation \cite{ncps}. Developed from first principles, LNNs differ in their neuron operation and are inspired by the adaptive behavior of biological neural systems, specifically the synaptic information transmission in Caenorhabditis elegans. This approach enables LNNs to emulate the flexibility and resilience of natural neural networks. By integrating a closed-form continuous-time approach, LNNs simulate ODE-based dynamics without relying on expensive ODE solvers, making them efficient to implement. Unlike static models, LNNs adapt continuously to new inputs, maintaining robustness in dynamic and noisy wireless environments \cite{robust_flight}. This adaptability makes LNNs particularly well-suited for real-world applications where channels varies along time. Additionally, LNNs can break down complex neural behaviors into interpretable patterns. They can enhance decision-making transparency and system resilience by utilizing techniques such as decision trees to analyze neural strategies, thereby further improving overall robustness \cite{Interpretability}.

\subsection{Comparison and Tradeoff Discussion}
The fundamental trade-off between these approaches centers on the balance between data dependency and interpretability.
Data-driven methods excel in dynamic, data-rich environments, offering superior performance with large datasets but are vulnerable to overfitting and lack transparency. These methods are often seen as black boxes, which can be problematic in applications requiring trust and verification. In contrast, model-driven methods are more interpretable, relying on well-established models backed by mathematical basis, and perform well even with limited data. However, they may struggle to adapt to unforeseen complexities, limiting their generalization to novel scenarios.
\par
Despite these differences, both approaches have their strengths and weaknesses, and the choice between them depends on the specific application requirements. Data-driven methods tend to shine in dynamic, data-rich environments where adaptability and performance are paramount. Model-driven methods, on the other hand, excel in environments where interpretability, low data requirements, and robustness to overfitting are more critical. A detailed comparison is provided in Table \ref{tab:tradeoff}. We will further explain how data-driven and model-driven methods differ in achieving robustness by discussing two specific examples in the next section.

\begin{figure*}[t]
    \centering
    \begin{subfigure}[b]{0.49\linewidth}
        \includegraphics[width=\linewidth]{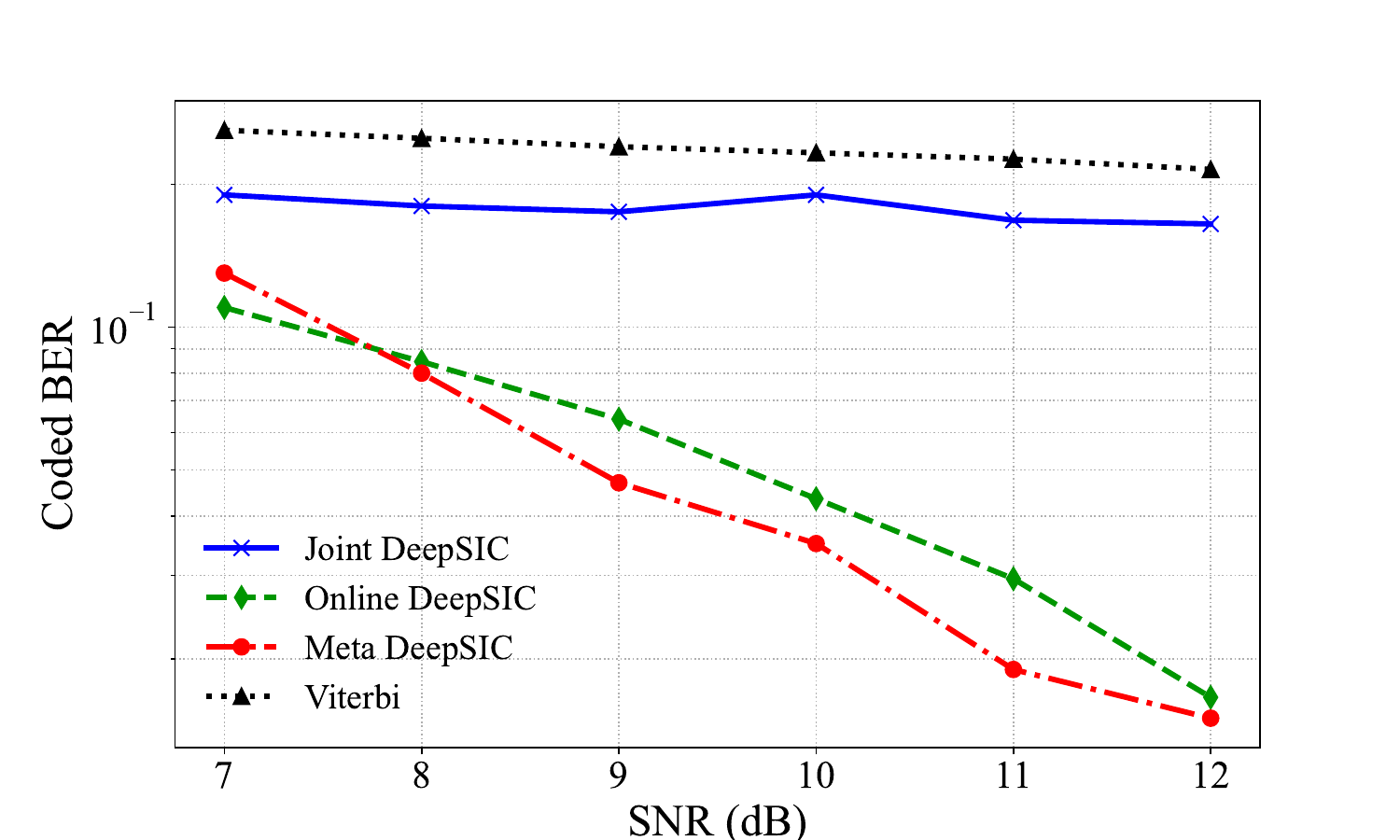}
        \caption{}
        \label{fig:BER}
    \end{subfigure}
    \hfill
    \begin{subfigure}[b]{0.47\linewidth}
        \includegraphics[width=\linewidth]{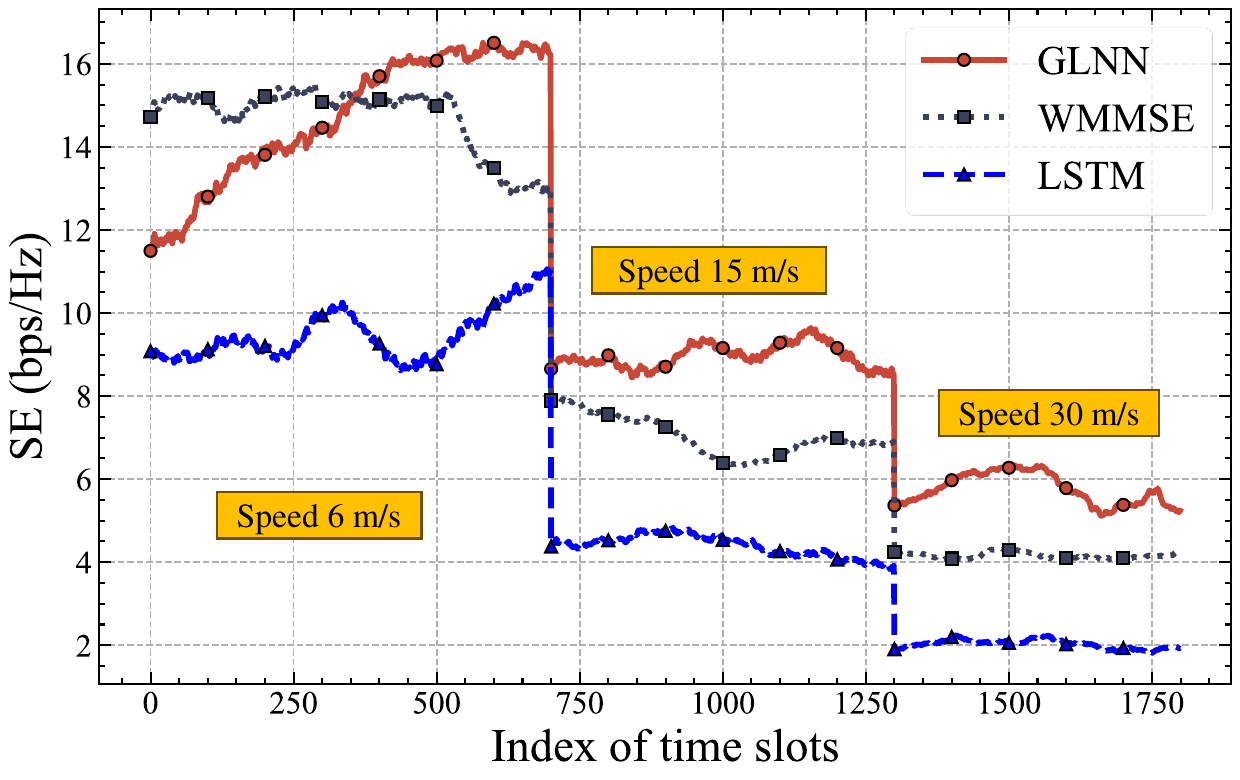}
        \caption{}
        \label{fig:dynamic_result}
    \end{subfigure}
    \captionsetup{font=footnotesize, name={Fig. }, labelsep=period}
    \caption{Performance Comparison of Robust Deep learning performance in physical layer communications. (a) Channel Decoding (BER vs SNR): Data-driven Meta-ViterbiNet achieves lowest BER, effectively handling non-linear channels.
(b) Beamforming (SE vs time): Model-driven, pretraining-free GLNN adapts quickly using minimal online data, achieving highest SE.}
    \label{fig:combined}
\end{figure*}

\section{Case Studies}\label{sec:cases}
In this section, we summarize the performance of DL-based robust schemes for physical layer communications by discussing two examples, one employing a data-driven method and the other a model-driven method, and how they addressed the robustness problem in broader 6G contexts.

\subsection{Channel Decoding with Data-Driven Method}\label{channel_decode}
We consider an end-to-end communication scenario over non-linear synthetic channels \cite{meta_deepsic}. The receiver employs the deep soft interference cancellation (DeepSIC) technology, where traditional decoding is replaced by DNNs. The traditional Viterbi equalizer is used as a baseline method. System parameters are detailed in Table \ref{sim_parameters_combined}. We compare the BER across joint training, online training, and meta-learning methods. Joint training involves training the DNN on a fixed dataset, while online training fine-tunes using data from the decoded blocks. The meta-learning approach utilizes the MAML algorithm, a data-driven meta learning method. Fig. \ref{fig:BER} shows the BER as a function of testing SNR for these methods, revealing that BER decreases with increasing SNR across all schemes. Notably, the inherent channel non-linearity causes mismatch for the Viterbi equalizer, limiting its effectiveness in this scenario. Conversely, the data-driven DeepSIC architecture can learn to compensate for these non-linear distortions. This data-driven meta-learning method consistently achieves the lowest BER when SNR exceeds 8dB, indicating its ability to handle the complex decoding task with non-linear and time-varying channels in previously unseen test environments.

\subsection{Beamforming with Model-Driven Method}\label{beamforming_ncps}
We consider a MIMO beamforming scenario in which a BS is equipped with 64 antennas and simultaneously serves 4 users, each equipped with 2 antennas \cite{ncps_beamforming}. The users move at varying speeds of 6 m/s, 15 m/s, and finally 30 m/s, with each phase comprising 700, 600, and 500 time slots, respectively. The detailed parameters are shown in Table \ref{sim_parameters_combined}. Fig. \ref{fig:dynamic_result} shows the average SE of the model-driven method in this scenario in comparison with other baselines. The pretraining-free gradient-based liquid neural network (GLNN) model rapidly surpasses the traditional WMMSE algorithm after a short adaptation period and consistently achieves higher SE compared to all other baselines, including the LSTM-based scheme. 
This strong performance underscores the advantages of the model-driven GLNN. The ability of GLNN to operate without pre-training and adapt effectively with little online data highlights the efficiency and robustness inherent in such model-driven designs, particularly beneficial in dynamic or data-scarce conditions.
\begin{table}[t]
\centering
\caption{
Simulation Parameters for Fig. \ref{fig:combined}}
\begin{tabular}{|c |c| c| c|} 
\cline{1-4}
\multicolumn{2}{|c|}{\textbf{Parameters (Fig. \ref{fig:BER})}} & \multicolumn{2}{c|}{\textbf{Parameters (Fig. \ref{fig:dynamic_result})}} \\ [0.2ex]
\cline{1-4}
Pilot Blocks & 300 & User Antenna Number & 2 \\
\cline{1-4}
Information Blocks & 300 & BS Antenna Spacing & $0.5 \, \lambda$ \\
\cline{1-4}
Block Symbols & 136 & BS Antenna Number & 64 \\
\cline{1-4}
Information Bits & 120 & Central Frequency & 28 GHz \\
\cline{1-4}
Iteration Blocks & 5 & User Number & 4 \\
\cline{1-4}
Power Attenuation & 0.5 & Transmit SNR & 10 dB \\
\cline{1-4}
Training SNR & 12 dB & Bandwidth (MHz) & 50 \\
\cline{1-4}
Max Epoch & 200 & Hidden Neuron Number & 30 \\
\cline{1-4}
\end{tabular}
\vspace{-4mm}
\label{sim_parameters_combined}
\end{table}
\subsection{Discussion on Generalization in Broader Contexts}\label{discussion}
The case studies demonstrate robust DL techniques applied to specific tasks. While illustrative, these examples highlight fundamental robustness principles applicable beyond these contexts. The data-driven meta-learning approach, showcased for channel decoding, demonstrates a powerful mechanism for rapid adaptation in changing environments. This capability is crucial for enhancing robustness in many other physical layer tasks pertinent to 6G, such as adaptive modulation and coding, link adaptation, or dynamic spectrum access. The model-driven GLNN used for beamforming highlights the potential for robust sequential tasks in dynamic system by incorporating structural priors. This is relevant for tasks beyond beamforming, including channel prediction, dynamic resource allocation, or interference management. These paradigms can be generalized for broader 6G contexts like ultra-reliable low-latency communications (URLLC). For URLLC, the rapid adaptation demonstrated by meta-learning could facilitate highly responsive resource allocation or scheduling policies that dynamically prioritize critical network traffic based on quickly learned environmental or traffic patterns.

\section{Future Research Directions}\label{sec:robust_challenges}
In this section, we discuss the potential future research directions for DL-based robust physical layer communications. The subsequent subsections explore specific areas where advancements are needed to fully realize the potential of DL in intelligent communications.

\subsection{Zero-Shot Learning}
Zero-shot learning (ZSL) refers to the ability of a model to recognize and classify new, unseen data classes without having been explicitly trained on them. This is critical for robust physical layer communications using DL, as wireless environments constantly evolve with new device types, interference patterns, or channel conditions (e.g., abrupt changes from line-of-sight to non-line-of-sight). Traditional DL often fails when encountering situations significantly different from its training data, such as attempting to recognize a novel modulation scheme or configuring appropriate beamforming patterns with unseen user distributions or mobility patterns.
\par
While DL methods have shown a degree of generalization capabilities, further exploration is needed to enhance their ability to adapt to new and unforeseen conditions without substantial performance loss. For instance, employing attribute-based learning, where models learn associations between signal properties like bandwidth, cyclostationary features, modulation characteristics and known classes, could allow generalization to new classes exhibiting novel combinations of these properties. Generative models like conditional generative adversarial networks could also facilitate ZSL. They can create synthetic signal samples for hypothesized, unobserved scenarios. If these synthesized samples effectively capture the characteristics of potential future conditions, then incorporating them into the knowledge of the model might allow it to bridge the gap and handle actual occurrences of these new conditions.

\subsection{Distributed Learning}
In large-scale wireless communication systems, deploying DL-based algorithms across multiple devices and network nodes is essential for enhancing scalability, fault tolerance, and resource efficiency.
However, applying DL in this domain faces significant challenges, including communication overhead, synchronization difficulties, statistical data heterogeneity and channel impairments. Therefore, a critical research direction is the development and configuration of distributed learning algorithms and frameworks specifically customized to particular application scenarios. For instance, in applications demanding ultra-low latency control, generic distributed methods may prove inadequate. Specific approaches for such scenarios involve designing algorithms with inherently faster convergence properties, adopting aggregation protocols like over-the-air computation that minimize communication delay by leveraging the physical layer for model aggregation, or employing asynchronous update mechanisms to improve synchronization efficiency. Beside, techniques like re-parametrization could be explored to help models better adapt to the unique local data distributions and conditions of each node. Furthermore, designing robust aggregation protocols, such as investigating alternatives to simple averaging (e.g., geometric median or other novel aggregation techniques) that are tolerant to channel impairments and node failures is vital for reliable communications.

\subsection{Multi-Modality Integration}
In wireless communication systems, integrating data from multiple modalities can significantly enhance the performance and reliability of physical layer algorithms. 
Instead of relying solely on the primary radio frequency (RF) signals used for communication, leveraging diverse information sources could provide a richer understanding of the communication scenario. These sources can include non-RF data such as user location or visual sensor inputs, as well as contextual RF information like local spectrum occupancy information or the detection of potential obstructions.
\par
For example, applications may include using data from visual sensors to proactively adjust modulation and coding schemes before a detected blockage impacts the RF link, or using combined positioning data and spectrum sensing data to perform more accurate beam tracking. Key challenges involve effectively fusing inherently heterogeneous data streams, which differ significantly in format, dimensionality, and structure, and developing efficient multi-modal DL structures without introducing prohibitive computational complexity.

\subsection{Cross-Layer Optimization and Integration}
To fully exploit the capabilities of DL in wireless communication systems, performing optimization across multiple network layers is essential. Cross-layer optimization can lead to significant performance improvements by ensuring that DL-informed decisions at different layers (e.g., physical, medium access control, network, application) are well-coordinated, achieving synergistic effects unobtainable through isolated layer operation. Future research should develop strategies for such cross-layer coordination leveraging DL. For instance, leveraging DL for physical layer channel prediction could facilitate proactive resource scheduling at the medium access control layer or enable dynamic routing decisions at the network layer ahead of significant channel quality shifts. Conversely, information from higher layers, such as application quality of service requirements or network congestion levels, could dynamically guide the controlling of physical layer parameters like beamforming and power allocation. Key challenges involve developing joint optimization techniques that appropriately balance trade-offs across layers (e.g., spectral efficiency versus end-to-end latency), and ensuring seamless integration and compatibility of these DL-driven cross-layer functions within existing network protocols and architectures.

\subsection{Future Standardization}
The incorporation of data-driven and model-driven models in physical layer communications may have implications for future standardization. Specifically, data-driven DL-based communication technologies will likely influence the collection, formatting, and storage of wireless data in communication systems. For instance, the use of data-driven models may facilitate the development of standardized protocols for data labeling, data quality control, and data compression, which can in turn enable more efficient and accurate data transmission. Moreover, the use of data-driven models may also impact the design of DL-based communications systems, including the development of new data management architectures and data analytic frameworks.
\par
Model-driven approaches are increasingly being integrated into communication system design to enhance performance and reliability. Due to the task-understanding ability of model-driven models, most of them exhibit improved tolerance to lost data or noise, which can lead to more resilient and efficient communication systems by reducing the need for excessive redundancy and error correction mechanisms. For instance, in beamforming applications, this could result in a relaxed accuracy requirement for CSI estimation and feedback. This is especially beneficial in the development of enhanced mobility management techniques and the reduction of redundancy in 6G radio access network standards \cite{3GPP_NR_Release18}.

\section{Conclusion}\label{sec:conclusion}
This article delved into the strategies and approaches for developing robust DL-based physical layer communications. Traditional DL models were reviewed with an examination of their characteristics and limitations. The discussion focused on both data-driven and model-driven robust schemes designed to enhance the robustness and adaptability of wireless communication systems. The discussion also included identifying current challenges and potential future research directions. Practical applications and effectiveness of both data-driven and model-driven methods were demonstrated through case studies. As wireless networks evolve, refining and advancing these DL-based technologies will be essential for maintaining robust, efficient, and adaptable 6G wireless communication systems.

\bibliographystyle{IEEEtran}
\bibliography{robust}
\vspace{12pt}
\end{document}